\documentclass[prd,onecolumn,nofootinbib,showpacs,showkeys]{revtex4}
\usepackage{bm,amsmath,amssymb}

\begin{document}

\title{Neutral-meson oscillations with torsion}
\author{Nikodem J. Pop{\l}awski}
\affiliation{Department of Physics, Indiana University, Swain Hall West, 727 East Third Street, Bloomington, Indiana 47405, USA}
\email{nipoplaw@indiana.edu}
\date{\today}

\begin{abstract}
We propose a simple mechanism that may explain the observed particle-antiparticle asymmetry in the Universe.
In the Einstein-Cartan-Sciama-Kibble theory of gravity, the intrinsic spin of matter generates spacetime torsion.
Classical Dirac fields in the presence of torsion obey the nonlinear Hehl-Datta equation which is asymmetric under a charge-conjugation transformation.
Accordingly, at extremely high densities that existed in the very early Universe, fermions have higher effective masses than antifermions.
As a result, a meson composed of a light quark and a heavy antiquark has a lower effective mass than its antiparticle.
Neutral-meson oscillations in thermal equilibrium therefore favor the production of light quarks and heavy antiquarks, which may be related to baryogenesis.
\end{abstract}


\maketitle

The Einstein-Cartan-Sciama-Kibble (ECSK) theory of gravity naturally extends general relativity (GR) to include the intrinsic angular momentum of matter \cite{KS,Hehl}.
In this theory, the spin of Dirac fields is a source of the torsion tensor $S^k_{\phantom{k}ij}$, which is the antisymmetric part of the affine connection: $S^k_{\phantom{k}ij}=\Gamma^{\,\,\,k}_{[i\,j]}$.
The Lagrangian density for a free Dirac spinor $\psi$ with mass $m$ is given by $\mathfrak{L}_\textrm{D}=\frac{i\sqrt{-g}}{2}(\bar{\psi}\gamma^i\psi_{;i}-\bar{\psi}_{;i}\gamma^i\psi)-m\sqrt{-g}\bar{\psi}\psi$, where $g$ is the determinant of the metric tensor $g_{ik}$ and the semicolon denotes a covariant derivative with respect to the affine connection $\Gamma^{\,\,k}_{i\,j}$.
We use units in which $\hbar=c=k_\textrm{B}=1$.
Varying $\mathfrak{L}_\textrm{D}$ with respect to the spinor adjoint conjugate $\bar{\psi}$ gives the Dirac equation $i\gamma^k(\psi_{:k}+\frac{1}{4}C_{ijk}\gamma^i\gamma^j\psi)=m\psi$, where the colon denotes a Riemannian covariant derivative (with respect to the Christoffel symbols) and $C_{ijk}=S_{ijk}+S_{jki}+S_{kji}$ is the contortion tensor \cite{Hehl,HD}.
Varying the total (gravity plus matter) Lagrangian density $-\frac{R\sqrt{-g}}{2\kappa}+\mathfrak{L}_\textrm{D}$ with respect to $C_{ijk}$ gives the Cartan relation \cite{KS} between the torsion tensor and the Dirac spin tensor $s^{ijk}=\frac{2}{\sqrt{-g}}\frac{\delta\mathfrak{L}_\textrm{D}}{\delta C_{ijk}}=\frac{1}{2}e^{ijkl}j_{\textrm{A}l}$, where the tensor $e^{ijkl}=\frac{\epsilon^{ijkl}}{\sqrt{-g}}$, $\epsilon^{ijkl}$ is the Levi-Civita permutation symbol and $j_{\textrm{A}}^k=\bar{\psi}\gamma^5\gamma^k\psi$ is the axial fermion current \cite{Hehl,HD}.
Substituting this quadratic (in spinor fields) relation to the Dirac equation gives the cubic Hehl-Datta equation for $\psi$ \cite{Hehl,HD}:
\begin{equation}
i\gamma^k\psi_{:k}=m\psi-\frac{3\kappa}{8}j_{\textrm{A}k}\gamma^5\gamma^k\psi.
\label{HeDa1}
\end{equation}
For a spinor with electric charge $e$ in the presence of the electromagnetic potential $A_k$, we must replace $\psi_{:k}$ by $\psi_{:k}-ieA_k\psi$:
\begin{equation}
i\gamma^k\psi_{:k}+eA_k\gamma^k\psi=m\psi-\frac{3\kappa}{8}(\bar{\psi}\gamma^5\gamma_k\psi)\gamma^5\gamma^k\psi.
\label{HeDa2}
\end{equation}

The charge conjugate $\psi^c$ of a spinor $\psi$ is defined as $\psi^c=-i\gamma^2\psi^\ast$ \cite{BD}.
The complex conjugate of (\ref{HeDa2}) leads to \cite{mat}
\begin{equation}
i\gamma^k\psi_{:k}^c-eA_k\gamma^k\psi^c=m\psi^c+\frac{3\kappa}{8}(\overline{\psi^c}\gamma^5\gamma_k\psi^c)\gamma^5\gamma^k\psi^c.
\label{HeDa3}
\end{equation}
Comparing (\ref{HeDa2}) with (\ref{HeDa3}) shows that $\psi$ and $\psi^c$ correspond to the opposite values of $e$, as expected from a charge-conjugation transformation.
They also satisfy different (classical) field equations because of the opposite signs of the corresponding Hehl-Datta cubic terms relative to the mass term.
Therefore, the classical Hehl-Datta equation leads to a charge-conjugation asymmetry between fermions and antifermions \cite{mat}.
This equation solved for fermion plane waves in the approximation of Riemann flatness gives the energy levels for a free fermion, $\omega=m+\epsilon$, where
\begin{equation}
\epsilon=\alpha\kappa N,
\label{eps}
\end{equation}
$N$ is the inverse normalization of the spinor's wave function, and $\alpha\sim1$ is a constant \cite{mat,Ker}.
These levels are higher than for the corresponding antifermion, $\omega=m-\epsilon$ \cite{mat}.
Since fermions have higher energy levels than antifermions due to the charge-conjugation asymmetric Hehl-Datta term, they are effectively more massive (in dispersion relations) and decay faster.
Such a difference between fermions and antifermions plays is significant only at extremely high densities where the contributions to the energy-momentum tensor from the spin density and from the energy density are on the same order \cite{Tra,non}.
In almost all physical situations, the ECSK gravity reduces to GR in which the Hehl-Datta term vanishes\footnote{
The Hehl-Datta equation corresponds to the axial-axial four-fermion interaction term in the Dirac Lagrangian.
If spinor fields, such as quarks, have a nonzero vacuum expectation value (form condensates), then this term can be the source of the observed small, positive cosmological constant \cite{dark}.
}
and the field equations are charge-conjugation symmetric.\footnote{
Torsion may also introduce an effective ultraviolet cutoff in quantum field theory for fermions \cite{non}.
The solutions of the nonlinear equations for the two-point function of a nonlinear spinor theory of \cite{Hei} exhibit self-regulation of its short-distance behavior \cite{Mer}.
The propagator of a similar theory based on the Hehl-Datta equation should thus also be self-regulated.
}

We now investigate how the above asymmetry of the Hehl-Datta equation affects the masses of mesons.
The effective masses due to torsion of a quark, $m_q$, and an antiquark, $m_{\bar{q}}$, are given by
\begin{eqnarray}
& & m_q=m_q^{(0)}+\epsilon, \label{cor1} \\
& & m_{\bar{q}}=m_{\bar{q}}^{(0)}-\epsilon,
\label{cor2}
\end{eqnarray}
where $m_q^{(0)}$ and $m_{\bar{q}}^{(0)}$ are the corresponding masses in GR, where the Hehl-Datta term is absent.
The sum of these masses is not affected by torsion, $m_q+m_{\bar{q}}=m_q^{(0)}+m_{\bar{q}}^{(0)}$.
The simplest toy relation for the effective mass of a ground-state meson, $m_{q\bar{q}}$, is
\begin{equation}
m_{q\bar{q}}=m_q+m_{\bar{q}}-\frac{m_q m_{\bar{q}}}{\mu},
\label{mes1}
\end{equation}
where the last term on the right-hand side is an interaction term with a negative sign due to the fact that a meson is a bound state of a quark and an antiquark.
The strength of an interaction between a quark and an antiquark is given by a mass-dimension parameter $\mu$.
Without torsion, the corresponding mass of a meson would be
\begin{equation}
m_{q\bar{q}}^{(0)}=m_q^{(0)}+m_{\bar{q}}^{(0)}-\frac{m_q^{(0)}m_{\bar{q}}^{(0)}}{\mu}.
\label{mes2}
\end{equation}
The relations (\ref{cor1}), (\ref{cor2}), (\ref{mes1}) and (\ref{mes2}) give
\begin{equation}
m_{q\bar{q}}=m_{q\bar{q}}^{(0)}+\frac{\epsilon}{\mu}(m_q^{(0)}-m_{\bar{q}}^{(0)}).
\end{equation}

If a meson is composed of a quark and its Dirac-adjoint antiquark, then $m_q^{(0)}=m_{\bar{q}}^{(0)}$.
In this case, the Hehl-Datta term does not affect the mass of such a meson, $m_{q\bar{q}}=m_{q\bar{q}}^{(0)}$.
As a result, the mass of the corresponding antimeson is the same as the mass of the meson.
If, however, a meson is composed of a quark and an antiquark with different flavors, then the mass of the corresponding antimeson is not equal to the mass of the meson.
For example, for neutral $B$ mesons,
\begin{eqnarray}
& m_{B_0}=m_{d\bar{b}}=m_{d\bar{b}}^{(0)}+\frac{\epsilon}{\mu}(m_d^{(0)}-m_{\bar{b}}^{(0)})=M_{B_0}-\Delta_{bd}, \label{bd1} \\
& m_{\bar{B}_0}=m_{b\bar{d}}=m_{b\bar{d}}^{(0)}+\frac{\epsilon}{\mu}(m_b^{(0)}-m_{\bar{d}}^{(0)})=M_{B_0}+\Delta_{bd},
\label{bd2}
\end{eqnarray}
where
\begin{equation}
M_{B_0}=m_{b\bar{d}}^{(0)}
\end{equation}
and
\begin{equation}
\Delta_{bd}=\frac{\epsilon}{\mu}(m_b^{(0)}-m_{\bar{d}}^{(0)})>0.
\label{delt}
\end{equation}
The relations (\ref{bd1}) and (\ref{bd2}) use $m_b^{(0)}=m_{\bar{b}}^{(0)}$, $m_d^{(0)}=m_{\bar{d}}^{(0)}$ and $m_{b\bar{d}}^{(0)}=m_{d\bar{b}}^{(0)}$.

Neutral-meson oscillations transform neutral particles with nonzero internal quantum numbers into their antiparticles.
For example, $B_0$ mesons transform into their antiparticles $\bar{B}_0$ through the weak interaction.
In thermal equilibrium at temperature $T$, the particle number densities $n$ of these mesons are related by
\begin{equation}
\frac{n_{\bar{B}_0}}{n_{B_0}}=e^{-(m_{\bar{B}_0}-m_{B_0})/T}=e^{-2\Delta_{bd}/T}.
\end{equation}
If $\Delta_{bd}\ll T$ then
\begin{equation}
\frac{n_{B_0}-n_{\bar{B}_0}}{n_{B_0}+n_{\bar{B}_0}}\approx \frac{\Delta_{bd}}{T}.
\label{nn}
\end{equation}
$B_0-\bar{B}_0$ oscillations in thermal equilibrium therefore produce more $d$ quarks than $\bar{d}$ antiquarks and more $\bar{b}$ antiquarks than $b$ quarks.
Generally, neutral-meson oscillations in thermal equilibrium favor the production of light quarks and heavy antiquarks.
Such an asymmetry caused by the fermion-torsion coupling may be related to baryogenesis \cite{mat}.

Since the inverse normalization $N$ of a Dirac spinor is on the order of the cube of its energy scale and such a scale in the early Universe is given by the temperature of the Universe, we have $N\sim T^3$ \cite{mat}.
Substituting this relation to (\ref{eps}), (\ref{delt}) and (\ref{nn}) gives
\begin{equation}
\alpha=\frac{n_{B_0}-n_{\bar{B}_0}}{n_{B_0}+n_{\bar{B}_0}}\approx \frac{\kappa(m_b^{(0)}-m_d^{(0)})T^2}{\mu},
\end{equation}
which then gives
\begin{equation}
T\approx m_{\textrm{Pl}}\biggl(\frac{\alpha\mu}{m_b^{(0)}-m_d^{(0)}}\biggr)^{1/2}\approx m_{\textrm{Pl}}\biggl(\frac{\alpha\mu}{m_b^{(0)}}\biggr)^{1/2}.
\end{equation}
Putting $m_b^{(0)}\approx 4\,\mbox{GeV}$, $\alpha\approx 10^{-10}$, which is on the order of the observed baryon-to-entropy ratio, and $\mu\approx 200\,\mbox{MeV}$, which is near the QCD scale parameter of the SU(3) gauge coupling constant, gives $T\approx 10^{-6}\,m_{\textrm{Pl}}\approx 10^{12}\,\mbox{GeV}$ ($m_{\textrm{Pl}}$ is the reduced Planck mass).
This value is near the freeze-out temperature found in \cite{mat}.

\end{document}